\documentclass{article}
\usepackage{amssymb}
\usepackage{graphicx}
\usepackage{amssymb}
\usepackage{amsmath} 
\usepackage{slashed}
\numberwithin{equation}{section}

\rmfamily

\newcommand\tr{\operatorname{tr}}

\newcommand\om{\omega}

\begin{document}

	\begin{center}
		\LARGE\bf\expandafter{'t Hooft-Polyakov monopole of higher generalized angular momenta}\\
	\end{center}
	\vskip0.5in
\begin{center}
\large\expandafter{Siu Fai, Hsu}
\vskip0.1in
\large\expandafter{Department of Mathematics
\vskip0.1in The University of Hong Kong
\vskip0.3in
June 2013}
\end{center}

\vskip0.3in

\abstract{We recall the quaternionic fomulation, which can simplify the computation of the linearized Yang-Mills-Higgs equation in the background of a 't Hooft-Polyakov monopole. We then study the solutions in the cases $j=0$, $j=1$ and $j\geq 2$ separately. In particular, we investigate the spectral properties of the monopoles. We focus on some of the bound states and show that as the generalized momentum increases, the $k-$th eigenvalue tends to 1. We show the existence of Feshbach resonance for $\om <1$ in the coupled system and calculated the partial cross section when $\om >1$.}

\section{Introduction}
Solitons, which are spatially localized solutions with finite energy
of non-linear differential equations, play an important role in various particle physics theories.
In the quantum field theories, perturbations around solitons are interpreted as particles. 
So studying perturbations amounts to studying particles under the influence of background soliton. 

One of the interesting theories is the $SU(2)$ Yang Mills theory.
The theory allows the existence of the monopole which explains the quantization of electric charge.
Perturbation of the theory in the BPS-limit has been carried out intensively and many interesting phenomenon is found. For example, these particles has infinitely many bound states. 

Recently, a new phenomenon of the monopoles is observed. In \cite{slowde}, they showed that considering the non-linear perturbation, these excitation of the monopoles decays very slowly. It motivates the work \cite{Hodge}, in which they considered the perturbations of BPS monopoles, perserving the hedgehog form of the monopole. They obtained a system of two coupled, differential equations. The two channels are interpreted as one massive channel and one massless channel. 
Upon evolution, the energy from the massive channel leaks slowly to the massless channel, which explains the slow decay of monopoles. 

Motivated by this work, \cite{main} provides a formalism to compute perturbations which does not necessarily preserve the hedgehog form. It is expected that the formalism can encode more information from the theory and allows us to discover new phenomenon. However, they have only computed the systems of equations when generalized angular momentum is 0. In this paper, the systems of equations corresponding to higher generalized angular momentum will be computed and the constraint due to the background gauge condition will be worked out as well, which is not worked out explicitly in \cite{main}.

The rest of the paper is organized as follows. In section 2, we review some of the basis of the Yang-Mills theory. In section 3, we recall the quaternionic formulation in \cite{main} and relate it to the perturbation of the Yang-Mills equation. In section 4, the linearized Yang-Mills equation will be decomposed into systems of equations. The $j=0$ sector will be recalculated and we will also calculate the $j=1$ and $j\geq 2$ sector. In section 5, we investigate some of the bound states and some coupled systems. We showed that higher generalized momentum sector still contains infinite bound states. Also, the $k-$eigenvalue tends to 1 as the generalized momentum increases. We also calculated the cross section of a system and showed that Feshbach resonance occurs for $\om < 1$. We also computed the cross section of this system for $\om > 1$. Finally, in section 6, we end with some observations and some future direction of research.

\section{Yang-Mills' theory}
\subsection{Yang-Mills' equation}

The $SU(2)$ Yang-Mills theory is determined by the Lagrangian density
\begin{equation}
{\cal L}= \frac{1}{2}\tr(D_\mu\phi)(D^\mu\phi)-\frac{1}{4}\tr F^{\mu\nu}F_{\mu\nu}-\lambda(|\vec{\phi}|^2-1)^2.
\end{equation}
Here $\phi=\phi^at_a$ is an $\mathfrak{su}(2)-$valued function on the (1+3)$-$dimensional spacetime and $A$ is an $SU(2)-$ connection, where $\left\{t_a\right\}$ is a basis of $\mathfrak{su}(2)$ satisfying $[t_a,t_b] = \epsilon_{abc}t_c$. The covariant derivative is given by $D_\mu\phi=\partial_\mu\phi+[A_\mu,\phi]$ and the field strength tensor of the connection $A_\mu=A_\mu^at_a$ is $F_{\mu\nu}=\partial_\mu A_\nu-\partial_\nu A_\mu+[A_\mu,A_\mu]$.

The equations of motion is 
\begin{subequations}\label{eqm}
\begin{align}
D_\mu D^\mu\phi = \lambda(1-|\phi|^2)\phi;\\
D_\mu F^{\mu\nu}=[D^\nu\phi,\phi].
\end{align}
\end{subequations}
The following three conserved quantities can be defined. The energy is 
\begin{equation*}E = \int \frac{1}{2}|D_\mu\phi|^2 + \frac{1}{4}|F_{\mu\nu}|^2 + \lambda(|\vec{\phi}|^2-1)^2.
\end{equation*}
The electric charge is 
\begin{equation*}Q = \int \tr D_i\phi E_i d^3x,\end{equation*}
and the magnetic charge is 
\begin{equation*}\Phi=\int\tr D_i\phi B_id^3x,\end{equation*}
where $E_i= F^{0i}$ and $B_i = \epsilon_{ijk}F_{jk}$.

Finiteness of the energy imposes a boundary condition on the solutions that $|\phi(\vec{x})|\to 1$ as $|\vec{x}|\to \infty$. So $\phi$ determines a map from $S^2_{\infty}$ to $S^2$. In fact, the degree of this map is proportional to the magnetic charge $\Phi$.

To minimize the energy, one can rearrange the terms in the integrand of the energy
\begin{equation} E = \int |D_0\phi|^2 + |E_i|^2 + |D_i\phi \pm B_i|^2 \mp \tr D_i\phi B_i d^3x.\end{equation}
This shows that $E\geq |\Phi|$ and the equality holds only when 
\begin{equation}\label{Bogo}
D_i\phi \pm B_i = 0
\end{equation} 
Equation (\ref{Bogo}) is called the Bogomol'nyi equation. The Bogomol'nyi equation implies the equation of motion (\ref{eqm}). From now on, we will concentrate on solutions of 
\begin{equation}\label{Bogo1}D_i\phi + B_i = 0
\end{equation}
If a solution of equation (\ref{eqm}) is in the hedgehog form 

\begin{equation}\label{hogform}
A_0=0,\; A^a_i = \epsilon_{aik}\frac{x^k}{r^2}(1-W(t,r)),\;\phi^a=\frac{x^a}{r^2}H(t,r),
\end{equation} 
then $W$ and $H$ satisfy 
\begin{subequations}\label{eqmHW}
\begin{align}
\Big(r^2\Big(\frac{\partial^2}{\partial t^2} - \frac{\partial^2}{\partial r^2}\Big)+W^2+H^2-1\Big)W=0;\\
\Big(r^2\Big(\frac{\partial^2}{\partial t^2} - \frac{\partial^2}{\partial r^2}\Big)+2W^2+\lambda(H^2-r^2)\Big)H=0.
\end{align}
\end{subequations}
Solutions in this form satisfies the Bogomol'nyi equation. If we let $\lambda=0$, which is called the BPS-limit, and assume that $W$ and $H$ does not depend on $t$, then equation~\ref{eqmHW} becomes
\begin{subequations}\label{eqmHW1}
\begin{align}
(-r^2\frac{\partial^2}{\partial r^2}+W^2+H^2-1)W=0;\\
(-r^2\frac{\partial^2}{\partial r^2}+2W^2)H=0.
\end{align}
\end{subequations}
In this case, there is indeed an analytic solution (see for example \cite{main2})

\begin{equation}\label{asol}
H(r)=1-r\coth r, W(r)=\frac{r}{\sinh r}.
\end{equation}
One can show that the corresponding magnetic charge is $2\pi$.

\subsection{Perturbation of the equations}
Now we turn to the perturbation of the theory. This means the linearization of the equations of motion or the Bogomol'nyi equation. We will perturb around the time-independent monopole solutions. 

By a monopole solution, it shall mean a time-independent solution $(\phi,A_\mu)$ of equation (\ref{Bogo1}) with $A_0=0$. Suppose $(\phi,A_i)$ is a monopole solution. We consider the pertubation around ($\phi, A_i)$, $\tilde{\phi}=\phi + e^{i\om t}\varphi, \tilde{A}_i=A_i+e^{i\om t}a_i$ so that $\tilde{\phi}$ satisfies equation (\ref{eqm}). Putting it into equation, we get
\begin{subequations}\label{leqm}
\begin{align}\label{maineq}
D_iD_i\varphi+[A_i,D_i\phi]+D_i[a_i,\phi]=-\om^2\varphi;\\
D_iD_ia_j-D_iD_ja_i+[a_i,F_{ij}]-[\phi,D_j\phi]-[\varphi,D_j\phi]-[\phi,[a_j,\phi]]=-\om^2a_j.
\end{align}
\end{subequations}
The rest of the paper will mainly focus on equation (\ref{maineq}).

We can also linearize the Bogomol'nyi equation in the same way, and get
\begin{equation}
\epsilon_{ijk}D_ja_k=D_i\varphi+[a_i,\phi].
\end{equation}

Note that since there is no time derivative in the Bogomol'nyi equation, the $e^{i\om t}$ factor on the $\varphi$ disappears. Solutions which are gauge equivalent are deemed to be physically equivalent. In terms of perturbation, we want to consider only perturbations that are orthogonal to infinitestimal transformation due to the gauge transformation $\tilde{\phi}= \phi+[\theta,\phi], \tilde{A}_i=A_i-D_i\phi$. That is 
\begin{equation}\int\tr (a_iD_i\theta)+\tr(\varphi [\theta,\phi])d^3x=0,\end{equation}
for all $\theta$ with compact support. After integration by parts, the we see the requirement is that
\begin{equation}\label{bggc} D_ia_i + [\phi,\varphi]=0.\end{equation}
It is called the background gauge condition.

\section{Quaternionic Formulation and Perturbation of the equations}

\subsection{Quaternionic Formulation}
In this section, we will recall the quaternionic formulation introduced in \cite{main}.

Since every solution $(\phi,A_\mu)$ of equation (\ref{eqm}) is gauge equivalent to a solution $(\tilde{\phi},\tilde{A}_\mu)$, where $\tilde{A}_0=0$, we can concentrate on the solutions in the moduli space 
\begin{center}${\cal M} = \left\{\mathrm{solutions\ of\ equation\ (\ref{eqm})\ in\ the\ form\ }(A_i,\phi)\right\}/\sim$,\end{center}
 where two solutions are equivalent if they are guage equivalent. For all $[(A_i,\phi)]\in{\cal M}$, we can naturally identify it with a $(\mathfrak{su}(2)\otimes \mathbb{H})-$valued function by writing it as ${\cal Q}=\phi + A_ie_i\colon \mathbb{R}^4\to \mathfrak{su}(2)\otimes \mathbb{H}$. We will show that it is a convenient choice.

Let $\mathbb{H}$ be the set of quaternions, and $\left\{1,e_1,e_2,e_3\right\}$ is a real basis of $\mathbb{H}$, where $1$ is the multiplicative identity and $\left\{e_i\right\}$ satisfies
\begin{equation}e_ie_j = -\delta_{ij}+\epsilon_{ijk}e_k.\end{equation}
The conjungation of ${\cal Q} = a + b_ie_i$ is given by 
\begin{equation*}\bar{\cal Q}= a - b_ie_i.\end{equation*}
We now define the Dirac-type operators on these functions. Let $[(\phi,A_i)]\in{\cal M}$ such that both $\phi$ and $A_i$ are independent of time. Define \cite{main} 

\begin{equation}
\slashed{D}=D_ie_i+[\phi,\cdot],\quad \slashed{D}^\dagger=D_ie_i-[\phi,\cdot].
\end{equation} 
Note that
\begin{equation}
\slashed{D}^\dagger\slashed{D} = -D^2_i-[\phi,[\phi,\cdot]]+e_i[D_i\phi-B_i,\cdot], \slashed{D}\slashed{D}^\dagger=-D^2_i-[\phi,[\phi,\cdot]]-e_i[D_i\phi+B_i,\cdot].
\end{equation}
If we further assume $(\phi,A_i)$ is a monopole solution, then \cite{main}
\begin{equation}
\slashed{D}^\dagger\slashed{D} = -D^2_i-[\phi,[\phi,\cdot]]+2e_i[D_i\phi,\cdot],
\slashed{D}\slashed{D}^\dagger=-D^2_i-[\phi,[\phi,\cdot]].
\end{equation}
The linearized equations can be expressed compactly by this formulation. Let $q = \varphi+a_ie_i$. Then
\begin{equation}
\slashed{D}\bar{q}=D_ia_i+[\phi,\varphi]+(D_i\varphi-\epsilon_{ijk}D_ja_k-[\phi,a_i])e_i.
\end{equation}
So the linearized background gauge condition (\ref{bggc}) means that the real part of $\slashed{D}\bar{q}$ is zero. Assuming this condition, then \cite{main}

\begin{align}
\nonumber\slashed{D}^\dagger\slashed{D}\bar{q}=&-(D_iD_i\varphi+[A_i,D_i\phi]+D_i[a_i,\phi])\\
&+ e_j(D_iD_ia_j-D_iD_ja_i+[a_i,F_{ij}]-[\phi,D_j\phi]-[\varphi,D_j\phi]-[\phi,[a_j,\phi]]).
\end{align}
That is, given the background gauge condition, $\slashed{D}^\dagger\slashed{D}\bar{q}$ is the left hand side of the linearized Yang-Mills equation (\ref{leqm}). Therefore, equation (\ref{leqm}) is just
\begin{equation}\label{qbareq}\slashed{D}^\dagger\slashed{D}\bar{q}=\om^2\bar{q}.\end{equation}
We can also consider the equation
\begin{equation}\label{peq}\slashed{D}\slashed{D}^\dagger p=\om^2 p.\end{equation}
Non-zero solutions of equation (\ref{qbareq}) are of one to one correspondance with these of equation (\ref{peq}), given by the map
\begin{equation}\slashed{D}\bar{q}=\om p, \slashed{D}^\dagger p=\om\bar{q}.\end{equation}
Equation (\ref{peq}) is often easier to investigate than equation (\ref{qbareq}). We will see that equation (\ref{peq}) in the following investigation is of simpler form. Furthermore, to require the background gauge condition, we can consider only those solutions $p$ whose real part is zero. 

\subsection{Structure of the wave equation}

If we further assume that the monopole solution $(\phi,A_i)$ is in the hedgehog form (\ref{hogform}), then
\begin{equation}\label{DBPS}\slashed{D}=(e_i\partial_i)+\frac{(1-W)}{r}(\vec{e}\times \hat{x})\cdot\vec{t}+\frac{H}{r}\hat{x}\cdot\vec{t},\end{equation}
where $t_i$ as an operator acts by adjoint representation of $\mathfrak{su}(2)$ and $\hat{x} = \frac{\vec{x}}{r}$. The term $e_i\partial_i$ can be written in the form
\begin{equation}e_i\partial_i=\hat{x}\cdot\vec{e}\partial_r+\frac{1}{r}\vec{e}\cdot(\hat{x}\times\vec{L}),
\end{equation}
where the orbital angular momentum operator is $L_i = -\epsilon_{ijk}x_j\partial_k$.
We can also compute
\begin{align}\label{DDBPS}
\nonumber\slashed{D}^\dagger\slashed{D} = &-\Delta-\frac{2(1-W)}{r^2}\vec{L}\cdot\vec{t}-\frac{(1-W)^2}{r^2}\vec{t}^2+\frac{(1-W)^2-H^2}{r^2}(\hat{x}\cdot\vec{t})^2\\&+\frac{2rH'-2(1+W)H}{r^2}(\hat{x}\cdot\vec{t})(\hat{x}\cdot\vec{e})+\frac{2WH}{r^2}(\vec{e}\cdot\vec{t}),
\end{align}
where Laplace operator can be expressed as
\begin{equation}\Delta=\frac{1}{r}\partial^2_rr+\frac{1}{r^2}\vec{L}^2.\end{equation}
Noting that $rH'-H=W^2-1$, the above can be simplied to 
\begin{align}\label{DDBPS1}
\nonumber\slashed{D}^\dagger\slashed{D} = &-\Delta-\frac{2(1-W)}{r^2}\vec{L}\cdot\vec{t}-\frac{(1-W)^2}{r^2}\vec{t}^2+\frac{(1-W)^2-H^2}{r^2}(\hat{x}\cdot\vec{t})^2\\
&+\frac{2(W^2-WH-1)}{r^2}(\hat{x}\cdot\vec{t})(\hat{x}\cdot\vec{e})+\frac{2WH}{r^2}(\vec{e}\cdot\vec{t}).
\end{align}
On the other hand,
\begin{equation}\label{DDBPS2}
\slashed{D}\slashed{D}^\dagger=-\Delta-\frac{2(1-W)}{r^2}\vec{L}\cdot\vec{t}-\frac{(1-W)^2}{r^2}\vec{t}^2+\frac{(1-W)^2-H^2}{r^2}(\hat{x}\cdot\vec{t})^2.
\end{equation}
Now we define the generalized angular momentum operator $\vec{J}$ to be 
\begin{equation}
\vec{J}=\vec{L}+\vec{s}+\vec{t},
\end{equation}
where $s_i$ is the operator which acts by communtator of $e_i$ on the quaternionic part of the function. Since $[J_i,J_j]=\epsilon_{ijk}J_k$, and $\vec{J}$ communtes with both $\slashed{D}$ and $\slashed{D}^\dagger$, we see that the solution of equations (\ref{qbareq}) and (\ref{peq}) can be decomposed into eigenfunctions of eigenvalues $-j(j+1)$ of $\vec{J}^2$. We will call those eigenfunctions of eigenvalues $-j(j+1)$ as the $j$ sector.

\section{Solving the wave equation}
\subsection{$j=0$ sector}
Since the operators $\vec{L}$, $\vec{s}$ and $\vec{t}$ commutes, we can decomposes eigenspaces of $\vec{J}$ into tensor product of eigenspaces of $\vec{L}$, $\vec{s}$ and $\vec{t}$. The isospin part is viewed as spin 1 representation and the quaternion part can be viewed as a direct sum of a spin 0 and a spin 1 representations. By the Clebsch-Gordan decomposition, the basis of $j=0$, $j=1$ and $j\geq 2$ sector consist of 4, 10 and 12 elements, respectively. The basis can be found easily following the decomposition rules. 

We will start to investigate solutions of the differential equations which have eigenvalue 0 of the operator $\vec{J}^2$. Its basis conists of \cite{main} 

\begin{align}
\nonumber v_1 &= \vec{e}\cdot\vec{t};\\\nonumber
v_2 &= \hat{x}\cdot\vec{t};\\\nonumber
v_3 &= \vec{e}\cdot(\hat{x}\times\vec{t});\\
v_4 &= (\hat{x}\cdot \vec{e})(\hat{x}\cdot\vec{t}),
\end{align}
where for example $v_1 = \vec{e}\cdot\vec{t} = t_1\otimes e_1 + t_2\otimes e_2 + t_3\otimes e_3$ is a constant function. Note that the operator $\vec{t}^2$ always have value $-2$ on these basis.

Let $\bar{q} = \sum^4_{k=1}\frac{1}{r}c_k(r)v_k$. Using equation (\ref{DDBPS1}) and Table~\ref{tablej0}, we get the following systems of equations. 

\begin{subequations}
\begin{align}
-c''_1+\frac{1}{r^2}((H^2+3W^2-2W+2WH-1)c_1+(-2W+2WH)c_4)=\om^2c_1,\\
-c''_4+\frac{1}{r^2}((2W^2+2W+2-2WH)c_4+(-W^2-H^2-2W+2WH+3)c_1)=\om^2c_4
\end{align}
\end{subequations}
and
\begin{subequations}
\begin{align}
-c''_2+\frac{1}{r^2}(2W^2c_2+4WHc_3)=\om^2c_2,\\
-c''_3+\frac{1}{r^2}((H^2+3W^2+1)c_3+2WHc_2)=\om^2c_3.
\end{align}
\end{subequations}
Letting $c_3=h$, $c_2=\sqrt{2}w$, we get
\begin{subequations}
\begin{align}
-\frac{d^2w}{dr^2}+\frac{2W^2}{r^2}w+\frac{2\sqrt{2}WH}{r^2}h=\om^2w\\
-\frac{d^2h}{dr^2}+\frac{H^2+3W^2-1}{r^2}h+\frac{2\sqrt{2}WH}{r^2}w=\om^2h,
\end{align}
\end{subequations}
which are the perturbation of the Hedgehog fields $H(r)$ and $W(r)$ being done in \cite{hodge}.
\begin{table}
\begin{center}
\caption{The action of the operators in (\ref{DBPS}) and (\ref{DDBPS1}) on the basis of $j=0$ sector}\label{tablej0}
\begin{tabular}{| c | c | c | c | c | c |}

\hline
 & $\vec{L}^2$ & $\vec{L}\cdot\vec{t}$ & $\hat{x}\cdot\vec{t}$ & $\hat{x}\cdot\vec{e}$ & $(\vec{e}\times\hat{x})\cdot\vec{t}$\\\hline
 $v_1=\vec{e}\cdot\vec{t}$ & 0 & 0 & $-v_3$ & $-v_2+v_3$ & $-2v_2+v_3$\\\hline
 $v_2=\hat{x}\cdot\vec{t}$ & $-2v_2$ & $2v_2$ & 0 & $v_4$ & $-v_1+v_4$\\\hline
 $v_3=\vec{e}\cdot(\hat{x}\times\vec{t})$ & $-2v_3$ & $v_3$ & $v_1-v_4$ & $-v_1+v_4$ & $2v_4$\\\hline
 $v_4=(\hat{x}\cdot\vec{e})(\hat{x}\cdot\vec{t})$ & $2v_1-6v_4$ & $-v_1+3v_4$ & 0 & $-v_2$ & $v_3$\\\hline
  & $\vec{e}\cdot\vec{t}$ & $(\hat{x}\cdot \vec{t})^2$ & $(\hat{x}\cdot\vec{e})(\hat{x}\cdot\vec{t})$ & $\vec{e}\cdot(\hat{x}\times\vec{L})$ & \\\hline
 $v_1=\vec{e}\cdot\vec{t}$ & $2v_1$ & $-v_1+v_4$ & $v_1-v_4$ & 0 & \\\hline
 $v_2=\hat{x}\cdot\vec{t}$ & $v_3$ & 0 & 0 & $v_1-v_4$ & \\\hline
 $v_3=\vec{e}\cdot(\hat{x}\times\vec{t})$ & $2v_2+v_3$ & $-v_3$ & $v_4$ & $-v_1-v_4$ & \\\hline
 $v_4=(\hat{x}\cdot\vec{e})(\hat{x}\cdot\vec{t})$ & $v_1-v_4$ & 0 & 0 & $-2v_2-v_3$ & \\\hline
\end{tabular}
\end{center}
\end{table}
We now define the solutions which satisfies the background gauge condition. The real part of $\slashed{D}\bar{q}$ is
\begin{equation}\label{bgeq1}(-b'_1-b'_4+\frac{1}{r}(-2b_4-2(1-W)b_1))v_2.\end{equation}
So the only requirement of the coefficients is that 
\begin{equation}
-c'_1-c'_4+\frac{1}{r}((2W-1)c_1-c_4)=0
\end{equation}
 and there is no further constraint on $c_2$ and $c_3$. In particular, no two different Hedgehog perturbations are gauge equivalent.

We turn to equation (\ref{DDBPS2}). Letting $p=\sum^4_{k=1}\frac{1}{r}c_k(r)v_k$, it again decomposes into a system of differential equations.

\begin{subequations}
\begin{align}
-c''_1+\frac{1}{r^2}((W^2-2W+1+H^2)c_1+2Wc_4)=\om^2c_1,\\
-c''_4+\frac{1}{r^2}(2(W^2+W+1)c_4+(W^2-2W+1-H^2)c_1)=\om^2c_4,
\end{align}
\end{subequations}

\begin{equation}
-c''_2+\frac{2W^2}{r^2}c_2=\om^2c_2,
\end{equation}
and
\begin{equation}\label{j03eq}
-c''_3+\frac{W^2+H^2+1}{r^2}c_3=\om^2c_3.
\end{equation}
We see the form of the equations are simplier. Moreover, background gauge condition in this case is simply $c_2=0$.

\subsection{$j=1$ sector}
The basis of the $j=1$ sector are components of the vectors
\begin{align}\nonumber
v_1&=\vec{t};\\\nonumber
v_2&=\hat{x}\times\vec{t};\\\nonumber
v_3&=\vec{e}\times\vec{t};\\\nonumber
v_4&=(\hat{x}\cdot\vec{e})\vec{t};\\\nonumber
v_5&=(\hat{x}\cdot\vec{t})\vec{e};\\\nonumber
v_6&=(\vec{e}\cdot\vec{t})\hat{x};\\\nonumber
v_7&=(\hat{x}\cdot\vec{t})\hat{x};\\\nonumber
v_8&=(\hat{x}\cdot\vec{e})(\hat{x}\times\vec{t});\\\nonumber
v_9&=(\hat{x}\cdot\vec{t})(\hat{x}\times\vec{e});\\
v_{10}&=(\hat{x}\cdot\vec{e})(\hat{x}\cdot\vec{t})\hat{x}.
\end{align}
Using the equality 
\begin{equation}\label{Ueq}
\hat{x}\cdot(\vec{e}\times\vec{t})\hat{x}=-(\vec{e}\cdot\hat{x})(\hat{x}\times\vec{t})+(\hat{x}\cdot\vec{t})(\hat{x}\times\vec{e})+\vec{e}\times\vec{t},
\end{equation}
we can compute the value of the operators on the basis in Table~\ref{tablej1}. 

Note that although those operators acts on the components of $v_i$s, they determine operators acting on the vectors. For example, $\vec{L}^2$ maps a component of $v_7$ to the corresponding component of $2v_1-6v_7$. So the coefficient of the three components of $v_i$ satisfies the same equation and we will not make a distinction between them.
\begin{table}
\begin{center}
\caption{The action of the operators in (\ref{DBPS}) and (\ref{DDBPS1}) on the basis of $j=1$ sector}\label{tablej1}
\begin{tabular}{| c | c | c | c | c | }
\hline
 & $\vec{L}^2$ & $\vec{L}\cdot\vec{t}$ & $\hat{x}\cdot\vec{t}$ & $\hat{x}\cdot\vec{e}$ \\\hline
$v_1=\vec{t}$ & 0 & 0 & $-v_2$ & $v_4$\\\hline
$v_2=\hat{x}\times\vec{t}$ & $-2v_2$ & $v_2$ & $v_1-v_7$ & $v_8$\\\hline
$v_3=\vec{e}\times\vec{t}$ & 0 & 0 & $v_4-v_6$ & $-v_2-v_5+v_6$\\\hline
$v_4=(\hat{x}\cdot\vec{e})\vec{t}$ & $-2v_4$ & $v_5-v_6$ & $-v_8$ & $-v_1$\\\hline
$v_5=(\hat{x}\cdot\vec{t})\vec{e}$ & $-2v_5$ & $2v_5$ & 0 & $-v_7-v_9$\\\hline
$v_6=(\vec{e}\cdot\vec{t})\hat{x}$ & $-2v_6$ & $-v_4+v_5$ & $v_3-v_8+v_9$ & $-v_3-v_7+v_8-v_9$\\\hline
$v_7=(\hat{x}\cdot\vec{t})\hat{x}$ & $2v_1-6v_7$ & $-v_1+3v_7$ & 0 & $v_{10}$\\\hline
$v_8=(\hat{x}\cdot\vec{e})(\hat{x}\times\vec{t})$ & $2v_3-6v_8$ & $v_8+v_9$ & $v_4-v_{10}$ & $-v_2$\\\hline
$v_9=(\hat{x}\cdot\vec{t})(\hat{x}\times\vec{e})$ & $-2v_3-6v_9$ & $v_3+3v_9$ & 0 & $v_5-v_{10}$\\\hline
$v_{10}=(\hat{x}\cdot\vec{e})(\hat{x}\cdot\vec{t})\hat{x}$ & $2v_4+2v_5+2v_6-12v_{10}$ & $-v_4-v_6+4v_{10}$ & 0 & $-v_7$\\\hline
& $\vec{e}\cdot(\hat{x}\times\vec{L})$ & $\vec{e}\cdot\vec{t}$ & $(\hat{x}\cdot\vec{t})^2$ & $(\hat{x}\cdot\vec{t})(\hat{x}\cdot\vec{e})$ \\\hline
$v_1=\vec{t}$ & 0 & $-v_3$ & $-v_1+v_7$ & $-v_8$\\\hline
$v_2=\hat{x}\times\vec{t}$ & $v_3-v_8$ & $v_4-v_5$ & $-v_2$ & $v_4-v_{10}$\\\hline
$v_3=\vec{e}\times\vec{t}$ & 0 & $-2v_1+v_3$ & $-v_3-v_9$ & $-v_1+v_3+v_7-v_8+v_9$\\\hline
$v_4=(\hat{x}\cdot\vec{e})\vec{t}$ & $-2v_2$ & $v_2-v_5+v_6$ & $-v_4+v_{10}$ & $v_2$\\\hline
$v_5=(\hat{x}\cdot\vec{t})\vec{e}$ & $-v_1+v_3+v_7-v_9$ & $-v_2-v_4+v_6$ & 0 & 0\\\hline
$v_6=(\vec{e}\cdot\vec{t})\hat{x}$ & $-v_1+v_7-v_8+v_9$ & $2v_6$ & $-v_6+v_{10}$ & $v_6-v_{10}$\\\hline
$v_7=(\hat{x}\cdot\vec{t})\hat{x}$ & $v_5+v_6-2v_{10}$ & $-v_3+v_8-v_9$ & 0 & 0\\\hline
$v_8=(\hat{x}\cdot\vec{e})(\hat{x}\times\vec{t})$ & $-2v_2+v_5-v_6$ & $-v_1+v_7-v_9$ & $-v_8$ & $-v_1+v_7$\\\hline
$v_9=(\hat{x}\cdot\vec{t})(\hat{x}\times\vec{e})$ & $-v_2-v_4+v_5+2v_{10}$ & $v_1-v_7-v_8$ & 0 & 0 \\\hline
$v_{10}=(\hat{x}\cdot\vec{e})(\hat{x}\cdot\vec{t})\hat{x}$ & $v_3-2v_7-v_8+2v_9$ & $v_6-v_{10}$ & 0 & 0\\\hline
& $(\vec{e}\times\hat{x})\cdot\vec{t}$ & & & \\\hline
$v_1=\vec{t}$ & $v_5-v_6$  & & & \\\hline
$v_2=\hat{x}\times\vec{t}$ & $v_9$  & & & \\\hline
$v_3=\vec{e}\times\vec{t}$ & $-v_2-v_4-v_5$ & & & \\\hline
$v_4=(\hat{x}\cdot\vec{e})\vec{t}$ & $-v_3+v_8$ & & & \\\hline
$v_5=(\hat{x}\cdot\vec{t})\vec{e}$ & $v_1-v_3-v_7-v_9$ & & & \\\hline 
$v_6=(\vec{e}\cdot\vec{t})\hat{x}$ & $-v_3-2v_7+v_8-v_9$ & & & \\\hline 
$v_7=(\hat{x}\cdot\vec{t})\hat{x}$ & $-v_6+v_{10}$ & & & \\\hline
$v_8=(\hat{x}\cdot\vec{e})(\hat{x}\times\vec{t})$ & $-v_5+v_{10}$ & & & \\\hline 
$v_9=(\hat{x}\cdot\vec{t})(\hat{x}\times\vec{e})$ & $v_2+v_4-v_{10}$ & & & \\\hline
$v_{10}=(\hat{x}\cdot\vec{e})(\hat{x}\cdot\vec{t})\hat{x}$ & $-v_3+v_8-v_9$ & & & \\\hline 
\end{tabular}
\end{center}
\end{table}
Let $\bar{q}=\sum^{10}_{k=1}\frac{1}{r}c_k(r)v_k$, substituting into equation (\ref{DDBPS1}), we get the following system of equations.

\begin{subequations}
\begin{align}
-c''_1&+\frac{1}{r^2}(((1-W)^2+H^2)c_1+(-2W^2-2WH+2)c_3-2Wc_7+(-2W^2+2)c_8+2WHc_9)=\om^2c_1,\\
-c''_3&+\frac{1}{r^2}((3W^2-2W-1+H^2)c_3-2WH(c_1+c_7)-2c_8+2Wc_9)=\om^2c_3,\\\nonumber
-c''_7&+\frac{1}{r^2}(2(W^2+W+1)c_7+((1-W)^2-H^2)c_1+2(W^2-WH-1)c_3+2(W^2-1)c_8-2WHc_9)\\&=\om^2c_7,\\
-c''_8&+\frac{1}{r^2}((W^2+5+H^2)c_8+2(-W^2+2WH+2)(c_1+c_3)+2WH(c_7-c_9))=\om^2c_8,\\\nonumber
-c''_9&+\frac{1}{r^2}(2(W^2+W+1)c_9+(W^2+2W-3-2WH+H^2)c_3-2WHc_7+(2W-2WH-2)c_8)\\&=\om^2c_9
\end{align}
\end{subequations}
and
\begin{subequations}
\begin{align}
-c''_2&+\frac{1}{r^2}((H^2+W^2+1)c_2+2(W^2-1)c_4-2WHc_5)=\om^2c_2,\\
-c''_4&+\frac{1}{r^2}((H^2+W^2-2W+3)c_4+2(W^2-1)c_2-2WHc_5+2(1-W)c_6-2Wc_{10})=\om^2c_4,\\
-c''_5&+\frac{1}{r^2}(2W^2c_5-2WHc_2+2(-1+W-WH)c_4-2(1-W)c_6-2c_{10})=\om^2c_5,\\
-c''_6&+\frac{1}{r^2}((3W^2-2W+1+H^2+2WH)c_6+2WH(c_4+c_5)-2W(1+H)c_{10})=\om^2c_6,\\
-c''_{10}&+\frac{1}{r^2}(2(W^2+2W+3-WH)c_{10}+2(-W^2+WH+1)c_2+((1-W)^2-H^2)c_4&\\\nonumber&+(-W^2+2W+3-H^2+2WH)c_6)=\om^2c_{10}.
\end{align}
\end{subequations}
We now consider the background gauge condition. The real part of $\slashed{D}\bar{q}$ is 
\begin{align}\nonumber
&(\frac{H}{r}b_2-b'_4-\frac{W}{r}c_5-\frac{1}{r}c_6)v_1+(-\frac{H}{r}b_1-b'_3-\frac{(1-W)}{r}b_3-b'_8-\frac{2}{r}b_8-\frac{W}{r}b_9)v_2\\&+(-\frac{H}{r}b_2-b'_5+\frac{W}{r}b_5-b'_6+\frac{2W-1}{r}c_6-b'_{10}-\frac{2}{r}b_{10})v_7,
\end{align}
where $b_k(r)=\frac{1}{r}c_k(r)$ for all $k$. So the constraints on the coefficients is that
\begin{subequations}
\begin{align}
-c'_4&+\frac{1}{r}(Hc_2+c_4-Wc_5-c_6)=0,\\
-c'_3&-c'_8+\frac{1}{r}(-Hc_1-Wc_3-c_8-Wc_9)=0,\\
-c'_5&-c'_6-c'_{10}+\frac{1}{r}(-Hc_2+(1+W)c_5+2Wc_6-c_{10})=0.
\end{align}
\end{subequations}
Now $p=\sum^{10}_{k=1}\frac{1}{r}c_k(r)v_k$, putting it into equation (\ref{DDBPS2}), we get
\begin{subequations}\label{j11}
\begin{align}
-c''_3&+\frac{1}{r^2}(((W-1)^2+H^2)c_3-2c_8+2Wc_9)=\om^2c_3,\\
\label{j11}-c''_8&+\frac{H^2+W^2+5}{r^2}c_8=\om^2c_8,\\
-c''_9&+\frac{1}{r^2}(2(W^2+W+1)c_9+(H^2-(1-W)^2)c_3+2(W-1)c_8)=\om^2c_9
\end{align}
\end{subequations}
and
\begin{subequations}\label{j12}
\begin{align}
-c''_4&+\frac{1}{r^2}((H^2+W^2-2W+3)c_4+2(1-W)c_6-2Wc_{10})=\om^2c_4,\\
-c''_5&+\frac{1}{r^2}(2W^2c_5-2(1-W)(c_4+c_6)-2c_{10})=\om^2c_5,\\
-c''_6&+\frac{1}{r^2}((H^2+W^2-2W+3)c_6+2(1-W)c_4-2Wc_{10})=\om^2c_6,\\
-c''_{10}&+\frac{1}{r^2}(2(W^2+2W+3)c_{10}+((1-W)^2-H^2)(c_4+c_6))=\om^2c_{10}.
\end{align}
\end{subequations}

The equations satisfied by $c_1,c_2$ and $c_7$ are not included here since the background gauge condition in this case reads that $c_1=c_2=c_7=0$ and these equations decoupled with the system of equations above.
\subsection{$j\geq2$ sector}

As we have seen above, the basis of $j=0$ sector are scalars, and the basis of $j=1$ sector are vectors. In principle, the basis of $j\geq 2$ sector all have $2j+1$ degree of freedom. To define the basis, we will define the following symemtric tensor product.

Given vectors $\vec{x}^1,...,\vec{x}^k$, we define a product $S(\vec{x}^1,\cdots,\vec{x}^k)$ which is a symmetric tensor product given by
\begin{align}
\nonumber S(\vec{x}^1\cdots\vec{x}^k)_{i_1,...,i_k}=\sum_{\sigma,\tau\in S_j}\sum^{[j/2]}_{l=0}&c^j_l(\vec{x}^{\tau(1)}\cdot\vec{x}^{\tau(2)})\cdots(\vec{x}^{\tau(2l-1)}\cdot x^{\tau(2l)})\delta_{i_{\sigma(1)}i_{\sigma(2)}}\cdots\delta_{i_{\sigma(2l-1)}i_{\sigma(2l)}}\\&\times x^{\tau(2l+1)}_{i_{\sigma(2l+1)}}\cdots x^{\tau(j)}_{i_{\sigma(j)}},
\end{align}
where the numbers $c^j_l$ is chosen so that $\sum_{i_1=i_2}S(\vec{x}^1\cdots\vec{x}^k)_{i_1,...,i_k}=0$ and $c_0=1$. This property, together with the fact that the product is symmetric, implies that the degree of freedom of this tensor is $2j+1$. To compute the $c^j_l$, we compute
\begin{align}
\nonumber &\sum_{i_1=i_2}S(\vec{x}^1\cdots\vec{x}^k)_{i_1\cdots i_k}\\\nonumber&=\sum^{[j/2]}_{l=1}(2^l(l!))^2((j-2l)!)(2j-l)c^j_l((\vec{x}^{1}\cdot\vec{x}^{2})\cdots(\vec{x}^{2l-1}\cdot \vec{x}^{2l})\delta_{i_3i_4}\cdots\delta_{i_{2l-1}i_{2l}}x^{2l+1}_{i_{2l+1}}\cdots x^{j}_{i_j}+\cdots)\\&+\sum^{[j/2]-1}_{l=0}2(2^l(l!))^2((j-2l)!)c^j_l((\vec{x}^{1}\cdot\vec{x}^{2})\cdots(\vec{x}^{2l+1}\cdot \vec{x}^{2l+2})\delta_{i_3i_4}\cdots\delta_{i_{2l+1}i_{2l+2}}x^{2l+3}_{i_{2l+3}}\cdots x^{j}_{i_j}+\cdots).
\end{align}
Comparing the terms, for $[j/2]>l\geq 0$, we get the recurrence relation
\begin{equation}
2(2^l(l!))^2((j-2l)!)c^j_l + (2^{l+1}(l+1)!)^2((j-2l-2)!)(2j-l-1)c^j_{l+1}=0,
\end{equation}
that is,
\begin{equation}
(j-2l)(j-2l-1)c^j_l+2(l+1)^2(2j-l-1)c^j_{l+1}=0.
\end{equation}
For example, $(\vec{x}^1\otimes\vec{x}^2)_{ij}=2(x^1_{i}x^2_{j}+x^1_{j}x^2_{i})-\frac{4}{3}(\vec{x}^1\cdot\vec{x}^2)\delta_{ij}$. The basis of the $j\geq 2$ sector is the components of the product of $j$ vectors
\begin{align}\nonumber
v_1&=S(\hat{x}^{\otimes j-2},\vec{e},\vec{t});\\\nonumber
v_2&=S(\hat{x}^{\otimes j-1},\vec{t});\\\nonumber
v_3&=S(\hat{x}^{\otimes j-1},\vec{e}\times\vec{t});\\\nonumber
v_4&=S(\hat{x}^{\otimes j-2},\hat{x}\times \vec{e},\vec{t});\\\nonumber
v_5&=S(\hat{x}^{\otimes j-1},\hat{x}\times \vec{t});\\\nonumber
v_6&=(\vec{e}\cdot\vec{t})S(\hat{x}^{\otimes j});\\\nonumber
v_7&=(\hat{x}\cdot\vec{t})S(\hat{x}^{\otimes j-1},\vec{e});\\\nonumber
v_8&=(\hat{x}\cdot\vec{e})S(\hat{x}^{\otimes j-1},\vec{t});\\\nonumber
v_9&=(\hat{x}\cdot\vec{t})S(\hat{x}^{\otimes j});\\\nonumber
v_{10}&=(\hat{x}\cdot\vec{t})S(\hat{x}^{\otimes j-1},\hat{x}\times\vec{e});\\\nonumber
v_{11}&=(\hat{x}\cdot\vec{e})S(\hat{x}^{\otimes j-1},\hat{x}\times\vec{t});\\
v_{12}&=(\hat{x}\cdot\vec{e})(\hat{x}\cdot\vec{t})S(\hat{x}^{\otimes j}).
\end{align}
Using the equalities
\begin{subequations}
\begin{align}
S(\hat{x},\vec{e}\times\vec{t})+S(\hat{x}\times\vec{e},\vec{t})-S(\hat{x}\times\vec{t},\vec{e})=0,\\
(\hat{x}\cdot\vec{t})S(\hat{x},\vec{e})+(\hat{x}\cdot\vec{e})S(\hat{x},\vec{t})=S(\hat{x}\times\hat{e},\hat{x}\times\vec{t})+S(\vec{e},\hat{t})+(\vec{e}\cdot\vec{t})S(\hat{x},\hat{x})
\end{align}
\end{subequations}
and equation (\ref{Ueq}), we can compute the value of the operators appeared in equation~\ref{DBPS} and~\ref{DDBPS1} on the basis in Table~\ref{tablej2}. As in the $j=1$ sector, it shall mean that the operators acting on a component of the tensor product will give the corresponding component of the resultant tensor product.
\begin{table}
\begin{center}
\caption{The action of the operators in (\ref{DBPS}) and (\ref{DDBPS1}) on the basis of $j=2$ sector}\label{tablej2}
\begin{tabular}{| c | c | c | c |}
\hline
 & $\vec{L}^2$ & $\vec{L}\cdot\vec{t}$ & $\hat{x}\cdot\vec{t}$ \\\hline
$v_1$ & $-(j-1)(j-2)v_1$ & $-(j-2)v_1$ & $-v_3-v_4$\\\hline
$v_2$ & $-j(j-1)v_2$ & $-(j-2)v_2$ & $-v_5$\\\hline
$v_3$ & $-j(j-1)v_3$ & $(j-1)v_4$ & $-v_6+v_8$\\\hline
$v_4$ & $-j(j-1)v_4$ & $v_3-(j-2)v_4$ & $v_1+v_6-v_7-v_8$\\\hline
$v_5$ & $-j(j+1)v_5$ & $v_5$ & $v_2-v_9$\\\hline
$v_6$ & $-j(j+1)v_6$ & $jv_7-jv_8$ & $v_3+v_{10}-v_{11}$\\\hline
$v_7$ & $2(j-1)v_1-j(j+1)v_7$ & $-(j-1)v_1+(j+1)v_7$ & 0\\\hline
$v_8$ & $2(j-1)v_1-j(j+1)v_8$ & $-v_6+v_7-(j-1)v_8$ & $-v_{11}$\\\hline
$v_9$ & $2jv_2-(j+1)(j+2)v_9$ & $-jv_2+(j+2)v_9$ & 0\\\hline
$v_{10}$ & $-2v_3+2(j-1)v_4-(j+1)(j+2)v_{10}$ & $v_3-(j-1)v_4+(j+2)v_{10}$ & 0\\\hline
$v_{11}$ & $2jv_3+2(j-1)v_4-(j+1)(j+2)v_{11}$ & $v_{10}+v_{11}$ & $v_8-v_{12}$\\\hline
$v_{12}$ & $2v_6+2jv_7+2jv_8-(j+2)(j+3)v_{12}$ & $-v_6-jv_8+(j+3)v_{12}$ & 0\\\hline
 & $(\hat{x}\cdot\vec{e})(\hat{x}\cdot\vec{t})$ & $(\hat{x}\cdot\vec{e})$ & $\vec{e}\cdot\vec{t}$\\\hline
$v_1$ & $-v_1+v_5-v_6+v_7+v_8$ & $-v_2-v_4$ & $-v_1$\\\hline
$v_2$ & $-v_{11}$ & $v_8$ & $-v_3$\\\hline
$v_3$ & $-v_2+v_3+v_9+v_{10}-v_{11}$ & $-v_5+v_6-v_7$ & $-2v_2+v_3$\\\hline
$v_4$ & $-v_3-v_4+v_{11}$ & $v_1-v_8$ & $v_2+v_3$\\\hline
$v_5$ & $v_8-v_{12}$ & $v_{11}$ & $-v_7+v_8$\\\hline
$v_6$ & $v_6-v_{12}$ & $-v_3-v_9-v_{10}+v_{11}$ & $2v_6$\\\hline
$v_7$ & 0 & $-v_9-v_{10}$ & $-v_5+v_6-v_8$\\\hline
$v_8$ & $v_5$ & $-v_2$ & $v_5+v_6-v_7$\\\hline
$v_9$ & 0 & $v_{12}$ & $-v_3-v_{10}+v_{11}$\\\hline
$v_{10}$ & 0 & $v_7-v_{12}$ & $v_2-v_9-v_{11}$\\\hline
$v_{11}$ & $-v_2+v_9$ & $-v_5$ & $-v_2+v_9-v_{10}$\\\hline
$v_{12}$ & 0 & $-v_9$ & $v_6-v_{12}$\\\hline
 & $\vec{e}\cdot(\hat{x}\times\vec{L})$ & $(\vec{e}\times\hat{x})\cdot\vec{t}$ & $(\hat{x}\cdot\vec{t})^2$\\\hline
$v_1$ & $(j-2)v_2+(j-2)v_4$ & $v_2+v_3$ & $-v_1+v_7$\\\hline
$v_2$ & $(j-1)v_1-(j-1)v_8$ & $-v_6+v_7$ & $-v_2+v_9$\\\hline
$v_3$ & $(j-1)(-v_1+v_5-v_6+v_7)$ & $-v_5-v_7-v_8$ & $-v_3-v_{10}$\\\hline
$v_4$ & $v_1+(j-1)v_8$ & $v_5+v_8$ & $-v_4+v_{10}$\\\hline
$v_5$ & $j(v_3-v_{11})+(j-1)v_4$ & $v_{10}$ & $-v_5$\\\hline
$v_6$ & $j(-v_2+v_9+v_{10}-v_{11})$ & $-v_3-2v_9-v_{10}+v_{11}$ & $-v_6+v_{12}$\\\hline
$v_7$ & $-v_2+v_3+j(v_9+v_{10})$ & $v_2-v_3-v_9-v_{10}$ & 0\\\hline
$v_8$ & $-2v_2+(j-1)v_4$ & $-v_3+v_{11}$ & $-v_8+v_{12}$\\\hline
$v_9$ & $v_6+jv_7-(j+1)v_{12}$ & $-v_6+v_{12}$ & 0\\\hline
$v_{10}$ & $(j+1)v_{12}+v_7-v_5-v_8$ & $v_5+v_8-v_{12}$ & 0\\\hline
$v_{11}$ & $(j-1)(v_1+v_8)+j(v_7-v_6)-2v_5$ & $-v_7+v_{12}$ & $-v_{11}$\\\hline
$v_{12}$ & $-2v_9+jv_{10}$ & $-v_3-v_{10}+v_{11}$ & 0\\\hline
\end{tabular}
\end{center}
\end{table}
Letting $\bar{q}=\sum^{12}_{k=1}\frac{1}{r}c_k(r)v_k$, substituting into equation (\ref{DDBPS1}), we get 
\begin{subequations}
\begin{align}
-c''_1&+\frac{1}{r^2}((-W^2-2(j-1)W+j^2-j+1+H^2)c_1-2(j-1)(Wc_7+c_8))=\om^2c_1,\\
-c''_5&+\frac{1}{r^2}((W^2+j^2+j-1+H^2)c_5+2(W^2-WH-1)c_1-2WHc_7+2(W^2-1)c_8)=\om^2c_5,\\\nonumber
-c''_6&+\frac{1}{r^2}((3W^2-2W+j^2+j-1+2WH+H^2)c_6+2(-W^2+WH+1)v_1+2WHc_7\\&+2(1-W+WH)c_8+2W(H-1)c_{12})=\om^2c_6,\\\nonumber
-c''_7&+\frac{1}{r^2}((2W^2+2(j-1)W+j^2-j)c_7+(3W^2-2W-1-2WH)c_1-2WHc_5\\&-2j(1-W)c_6+2(W-WH-1)c_8-2jc_{12})=\om^2c_7,\\\nonumber
-c''_8&+\frac{1}{r^2}((W^2-2jW+j^2+3j-1+H^2)c_8+2WH(c_1-c_7)+2(W^2-1)c_5\\&+2j(1-W)c_6-2jWc_{12})=\om^2c_8,\\\nonumber
-c''_{12}&+\frac{1}{r^2}((2W^2+2(j+1)W+j^2+3j+2-2WH)c_{12}+(-2W^2+2WH+2)c_5\\&+(-W^2-2W+3+2WH-H^2)c_6+(H^2-(1-W)^2)c_8=\om^2c_{12}
\end{align}
\end{subequations}
and
\begin{subequations}
\begin{align}
\nonumber-c''_2&+\frac{1}{r^2}((W^2-2jW+j^2+j-1+H^2)c_2+(-2W^2-2WH+2)c_3+2WH(c_4+c_{10})\\&-2jWc_9+(2-2W^2)c_{11})=\om^2c_2,\\\nonumber
-c''_3&+\frac{1}{r^2}((3W^2-2W+j^2-j-1+H^2)c_3-2WH(c_2+c_9)+(-2W^2+2W+4WH)c_4+2Wc_{10}\\&+(2-2W^2)c_{11})=\om^2c_3,\\\nonumber
-c''_4&+\frac{1}{r^2}((-W^2-2(j-1)W+j^2+j-1+H^2+2WH)c_4-2(j-1)(1-W)c_3\\&-2(j-1)(Wc_{10}+c_{11}))=\om^2c_4,\\\nonumber
-c''_9&+\frac{1}{r^2}((2W^2+2jW+j^2+j)c_9+((1-W)^2-H^2)c_2+(2W^2-2WH-2)c_3\\&-2WHc_{10}+(2W^2-2)c_{11}=\om^2c_9,\\\nonumber
-c''_{10}&+\frac{1}{r^2}((2W^2+2jW+j^2+j)c_{10}+(W^2+2W-3+H^2-2WH)c_3+((1-W)^2-H^2)c_4\\&-2WHc_9+(2W-2-2WH)c_{11}=\om^2c_{10},\\\nonumber
-c''_{11}&+\frac{1}{r^2}((W^2+j^2+3j+1+H^2)c_{11}+(-2W^2+2WH+2)(c_2+c_3-c_4)+2WH(c_9-c_{10})\\&=\om^2c_{11}.
\end{align}
\end{subequations}

We turn to the background gauge condition. By letting $b_k=\frac{1}{r}c_k$, subtituting into equation (\ref{DBPS}), we see that the real part of $\slashed{D}\bar{q}$ is
\begin{align}\nonumber
&(-b'_1-b'_8+\frac{1}{r}((j-1-W)b_1+Hb_5-jb_6-Wb_7-2b_8))v_2\\\nonumber &+ (-b'_3-b'_{11}+\frac{1}{r}(-Hb_2+(1-W)(-b_3+b_4)+(j-1)b_5-Wb_{10}-2b_{11}))v_5\\
&+(-b'_6-b'_7-b'_{12}+\frac{1}{r}(-Hb_5+(j-2+2W)b_6+(j-1+W)b_7-2b_{12}))v_9.
\end{align}
So the constraint on the coefficient is
\begin{subequations}
\begin{align}
-c'_1&-c'_8+\frac{1}{r}((j-W)c_1+Hc_5-jc_6-Wc_7-c_8)=0,\\
-c'_3&-c'_{11}+\frac{1}{r}(-Hc_2+Wc_3+(1-W)c_4+(j-1)c_5-Wc_{10}-c_{11})=0,\\
-c'_6&-c'_7-c'_{12}+\frac{1}{r}(-Hc_5+(j-1+2W)c_6+(j+W)c_7-c_{12})=0.
\end{align}
\end{subequations}
Let $p=\sum^{12}_{k=1}\frac{1}{r}c_k(r)v_k$, putting $p$ into equation (\ref{DDBPS2}), we obtain

\begin{subequations}
\begin{align}\label{j21}
-c''_3&+\frac{1}{r^2}((H^2+W^2-2W+j^2-j+1)c_3+(2W-2)c_4-2Wc_{10}-2jc_{11})=\om^2c_3,\\\nonumber
-c''_4&+\frac{1}{r^2}((H^2+W^2+(2-2j)W+j^2+j-3)c_4+(2j-2)(W-1)c_3\\&-(2j-2)Wc_{10}-2(j-1)c_{11})=\om^2c_4,\\
-c''_{10}&+\frac{1}{r^2}((2W^2+2jW+j^2+j)c_{10}+(H^2-(1-W)^2)(c_3-c_4)-2(1-W)c_{11}=\om^2c_{10},\\
\label{j21d}-c''_{11}&+\frac{1}{r^2}(H^2+W^2+j^2+3j+1)c_{11}=\om^2c_{11}
\end{align}
\end{subequations}
and
\begin{subequations}\label{j22}
\begin{align}
-c''_1&+\frac{1}{r^2}((H^2+W^2+(2-2j)W+j^2-j-1)c_1+(2-2j)Wc_7+(2-2j)c_8)=\om^2c_1,\\
-c''_6&+\frac{1}{r^2}((H^2+W^2-2W+j^2+j+1)c_6+(2-2W)c_8-2Wc_{12})=\om^2c_6,\\\nonumber
-c''_7&+\frac{1}{r^2}((2W^2+(2j-2)W+j^2-j)c_7+((W-1)^2-H^2)c_1-2j(1-W)c_6\\&-2(1-W)c_8-2jc_{12})=\om^2c_7,\\
-c''_8&+\frac{1}{r^2}((H^2+W^2-2jW+j^2+3j-1)c_8+(2j-2jW)c_6-2jWc_{12}=\om^2c_8,\\
-c''_{12}&+\frac{1}{r^2}((2W^2+(2j+2)W+j^2+3j+2)c_{12}+((W-1)^2-H^2)(c_6+c_8)=\om^2c_{12}.
\end{align}
\end{subequations}
The above systems of equations have not included the coefficients $c_2, c_5$ and $c_9$ since the background gauge condition now reads $c_2=c_5=c_9=0$.
\section{Spectral properties of monopoles}
\subsection{Bound states}
In this section, we will use the analytic solution (\ref{asol}), although the procedure can be carried out for other solutions to equation (\ref{eqmHW1}).
It is well-known that the moduli space of monopoles is a disjoint union of $4k-$dimensional manifolds ${\cal M}_k$ \cite{godbook}. There is a natural action of $\mathbb{R}^3\times S^1$ on ${\cal M}_k$, where the $\mathbb{R}^3$ acts by translation of the origin and the $S^1$ is generated by the infinitestmal transformation
\begin{equation}\delta\phi = 0, \delta A_i= D_i\phi.\end{equation} 
In terms of the quaternionic formulation, it is just
\begin{equation}q_0 = (D_j\phi)e_j.\end{equation}
The generator of the translation can also be written in the form 
\begin{equation}q_i = e_i(D_j\phi e_j).\end{equation}
One can check that $\slashed{D}q_\mu = 0$ for $\mu=0,1,2,3$ and so they are solutions of equation (\ref{qbareq}) which satisfies the background gauge condition with $\omega = 0$.

\cite{bstate} in particular investigates equation (\ref{j03eq}) and found that there are infinite number of bound states of $\omega < 1$. We will consider the general case for higher $j$. Together with equation (\ref{j11}) and (\ref{j21d}), we see that solution of 
\begin{equation}\label{bseq} -\beta''+ \frac{H^2+W^2+j^2+3j+1}{r^2}\beta = \om^2\beta \end{equation}
is a solution to equation (\ref{peq}) for all $j\geq 0$. 

Here we will use numerical program \cite{prog} for Sturn-Lioville type problems to solve for the eigenvalues.
\begin{table}
\begin{center}
\caption{Eigenvalues of equation (\ref{bseq}) of different $j$}\label{evj}
\begin{tabular}{| c | c | c | c | c | c | c |}\hline
 & $\om^2_1$ & $\om^2_2$ & $\om^2_3$ & $\om^2_5$ & $\om^2_{10}$ & $\om^2_{100}$\\\hline
$j=0$ & 0.76882 & 0.89492 & 0.94012 & 0.97302 & 0.99187 & 0.99986\\\hline
$j=1$ & 0.88926 & 0.93771 & 0.96012 & 0.97964 & 0.99307 & 0.99986\\\hline
$j=2$ & 0.93750 & 0.96000 & 0.97222 & 0.98438 & 0.99408 & 0.99987\\\hline
$j=5$ & 0.97959 & 0.98438 & 0.98765 & 0.99174 & 0.99609 & 0.99989\\\hline
$j=50$ & 0.99972 & 0.99973 & 0.99974 & 0.99976 & 0.99979 &0.99995\\\hline
\end{tabular}
\\p.s. $\om^2_k$ represents the $k-$th eigenvalue.
\end{center}
\end{table}
Since 
\begin{equation}\frac{W^2+H^2+j^2+3j+1}{r^2}\approx 1-\frac{2}{r}+\frac{j^2+3j+2}{r^2}+O(e^{-r})\mathrm{\ as\ } r\to\infty,
\end{equation}
the potential is approximately Comlomb's potential with $l=j+1$, so the regular wavefunctions corresponding to the $k-$th eigenvalues for $j$ has the asymptotic behaviour
\begin{equation} u(r) \to r^{k+j+1}e^{-r/(k+j+1)},\mathrm{\ as\ }r\to\infty.\end{equation}
The corresponding eigenvalues can be approximated by
\begin{equation}\label{enapp} \om^2_k = 1 - \frac{1}{(k+j+1)^2},\end{equation}
which tends to 1 as $k$ or $j$ tends to infinity. It also shows that we should expect that we have infinite number of bound states. \cite{bstate} has shown that it is indeed the case for $j=0$. We computed the eigenvalues corresponding to some of the values of $k$ and $j$ in table (\ref{evj}). It agrees with our prediction and the values are indeed very close the approximation (\ref{enapp}).
\subsection{Scattering}
In this subsection, we consider the scattering problem which is in the form
\begin{equation}\label{scaeq}\Big(-\frac{d^2}{dr^2}+\frac{l(l+1)}{r^2}+V(r)-\om^2\Big)u=0,\end{equation}
where $V(r)\to 0$ as $r\to\infty$ exponentially. Asymptotically,
\begin{equation}\label{bessel}u(r)\approx r(A_l(\om)j_l(\om r)+B_l(\om)y_l(\om r)),\mathrm{\ for\ }r\mathrm{\ large},\end{equation}
where $j_l(r)$ and $B_l(r)$ are spherical Bessel functions and 
\begin{equation}j_l(\om r)\approx\sin\Big(\om r-\frac{l}{2}\pi\Big)/(\om r), y_l(\om r)\approx-\cos\Big(\om r-\frac{l}{r}\pi\Big)/(\om r).\end{equation}
So 
\begin{equation}u(r)\approx\frac{\sqrt{A^2_l(\om)+B^2_l(\om)}}{\om^2}\sin(\om r+\delta_l(\om)-\frac{l}{2}),\end{equation}
where $\delta_l(\om)=-\tan^{-1}\frac{B_l(\om)}{A_l(\om)}$ is the phase shift. The partial cross section is given by 
\begin{equation}\sigma_l(\om) = \frac{4\pi}{\om^2}(2l+1)\sin^2(\delta_l(\om)),\end{equation}
and the total cross section is 
\begin{equation}\sigma(\om)=\sum^{\infty}_{l=0}\sigma_l(\om)=\frac{4\pi}{\om^2}\sum^\infty_{l=0}(2l+1)\sin^2(\delta_l(\om)).\end{equation}
Back to our equations, consider equation (\ref{j22}), by letting $c_6=c_8=c_{12}=0$, we get
\begin{subequations}\label{syseq}
\begin{align}
-\beta'' &+ \frac{H^2+W^2+j^2-j-1}{r^2}\beta+\frac{(2-2j)W}{r^2}\alpha = \om^2\beta\\
-\alpha''&+ \frac{2W^2+j^2-j}{r^2}\alpha-\frac{2jW}{r^2}\beta=\om^2\alpha,
\end{align}
\end{subequations}
for all $j\geq 2$, where $\alpha=c_1$, $\beta=c_1+c_7$. If we let the decoupling constant to be 0, we get
\begin{subequations}
\begin{align}
\label{syseq1}-\beta'' &+ \frac{H^2+W^2+j^2-j-1}{r^2}\beta = \om^2\beta\\
-\alpha''&+ \frac{2W^2+j^2-j}{r^2}\alpha=\om^2\alpha,
\end{align}
\end{subequations}
The first equation is just equation (\ref{bseq}) after the substitution of $j$ by $j+2$ and the second equation is the scattering problem equation (\ref{scaeq}), with $V(r) = \frac{2W^2}{r^2}$ and $l=j-1$.

Back to equation (\ref{syseq}). For $\om < 1$, we can solve numerically for the regular solution for each $\om$ by imposing the condition that $\beta \approx Ar^2$ as $r\to 0$, $\beta(\infty) = 0$ and $\alpha \approx r^2$ as $r\to 0$ \cite{Hodge}. In this case, the solution satisfies
\begin{equation}\label{abapp}\beta(r) \to C(\om)r^{1/\nu(\om)}e^{-\nu r}, \alpha(r)\to D(\om)\sin\Big(\om r -\frac{(j-1)\pi}{2} + \delta(\om)\Big),\mathrm{\ as\ }r\to\infty,\end{equation}
for some coefficient $C(\om)$ and $D(\om)$, where $\nu(\om) = \sqrt{1-\om^2}$.

Comparing the solution of $\alpha$ with equation (\ref{bessel}) at some values of $r$, we compute the phase shift $\delta(\omega)=\delta_{j-1}$ and therefore the cross section $\sin^2(\delta(\omega))$.
\begin{figure}
\begin{center}
\includegraphics[scale = 0.5]{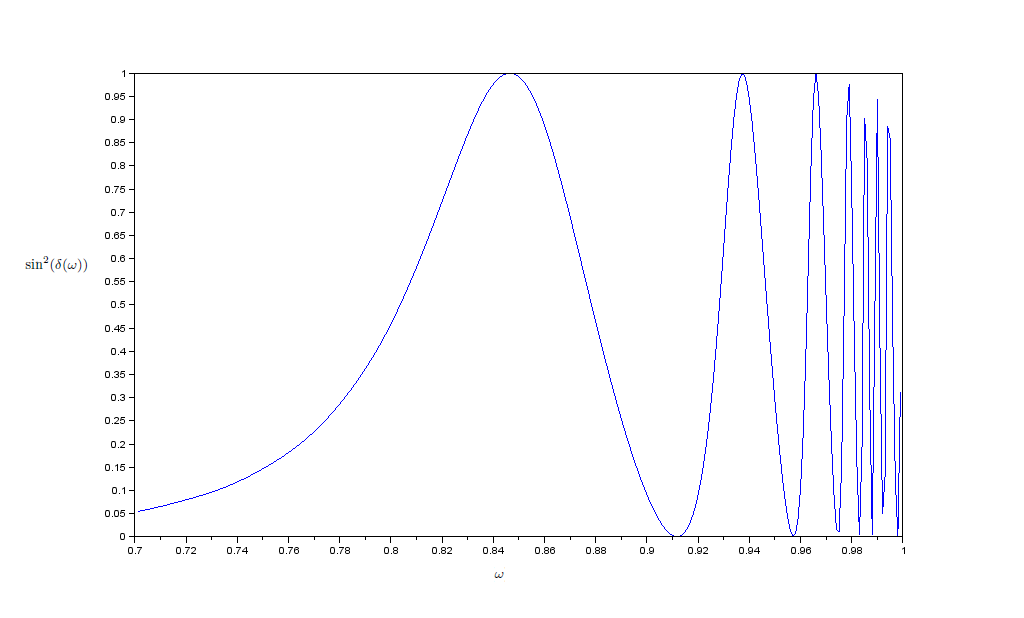}
\caption{For the system (\ref{syseq}) with $j=2$, the partial scattering cross section $\sin^2\delta(\omega)$ is plotted against $\omega$.}
\end{center}
\end{figure}

From the Figure 1, we see that the peak of $\sin(\delta(\omega))$ is close to the eigenvalues of equation (\ref{syseq1}), which is shown in the first row of the table~\ref{evj}. It shows the existence of the Feshbach resonance. 

The Feshbach resonance \cite{fesv} occurs when two channels are coupled and the energy of a discrete bound state of one channel lies within the continumm spectrum of another channel. In this case, when the energy of the incoming free particle is close to the energy of the bound state, the particle may decay to the bound state as an intermediate stage before being scattered out. The scattering length will become infinite as the energy tends to that of the bound state. It has been confirmed experimentally \cite{be} that we can tone the scattering length as large as possible by fine toning the energy closer to that of the bound state.

We also plotted the graph of the wavefunction at the $\om$ correspond to the peak of $\sin^2\delta(\om)$, where we denote by $\om_k$ the energy corresponding to the $k-$th peak.
\begin{figure}
\begin{center}
\includegraphics[scale = 0.5]{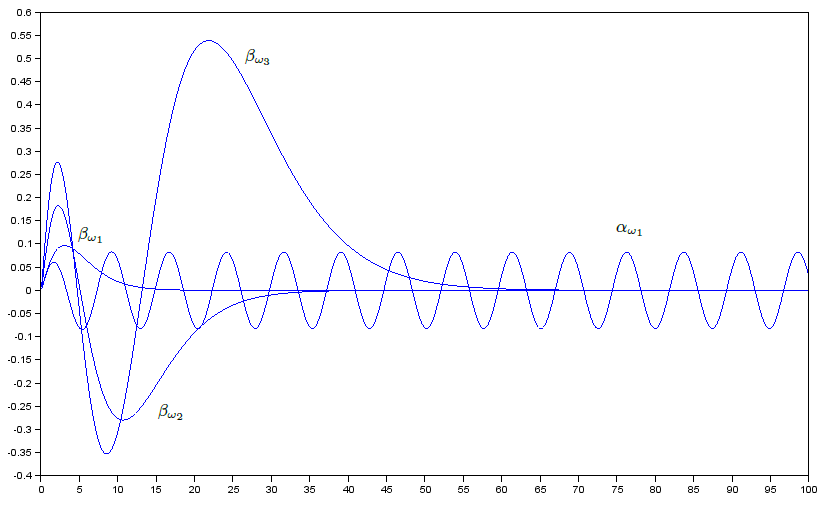}\\
\caption{$\alpha$ against $r$ at $\omega_1$, and $\beta$ against $r$ at $\omega_1$, $\om_2$ and $\om_3$.}
\end{center}
\end{figure}

We see that $\alpha\approx \sin(\om r + \delta(\om) - \frac{\pi}{2})$, while $\beta$ corresponding to $\omega_n$ oscillates $n/2-$cycles before dying down as expected in equation (\ref{abapp}).

For $\om > 1$, we can solve the system (\ref{syseq}) numerically by imposing regularity conditions, the solution should satisfies \cite{hsc}
\begin{equation}\label{abapp1}\beta(r) \to C(\om)\sin\Big(kr-\frac{(j-1)\pi}{2}-\eta\ln 2kr + \delta'(\om)\Big), \alpha(r)\to D(\om)\sin\Big(\om r -\frac{(j-1)\pi}{2} + \delta(\om)\Big),\mathrm{\ as\ }r\to\infty,\end{equation}
where $k = \sqrt{\om^2-1}$ and $\eta = -1/k$. We solved the case $j=2$ and computed the partial cross section $\sin^2\delta(\om)$ and $\sin^2\delta'(\om)$ for each $\om > 1$. The result is plotted in figure 3.
\begin{figure}
\begin{center}
\includegraphics[scale = 0.5]{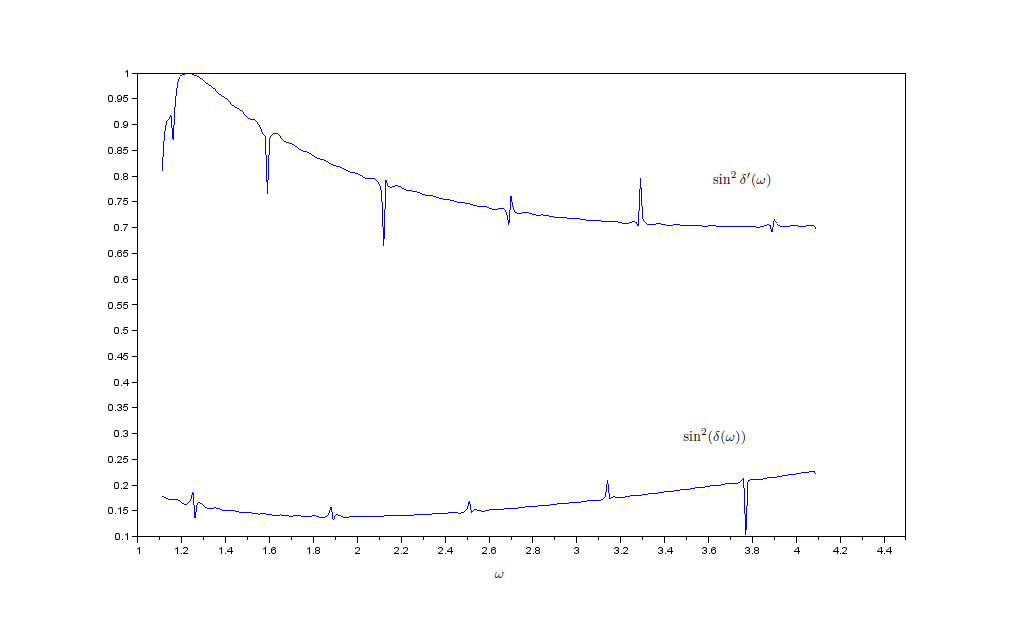}
\caption{$\sin^2\delta$ and $\sin^2\delta'$ against $\om >1$ for $j=2$.}
\end{center}
\end{figure}
From the figure, we predict that the both cross section approaches a fixed limit when $\om\to\infty$. The graphs are generally smooth except there are certain fluctuations on them. The fluctuation of one channel may cause that of another channel with certain delay.
\section{Conclusion}
We end with some observations. In this paper, we only considered the spherical symmetric perturbations. It is evident that these perturbations occur as a result of partial wave analysis of general perturbations. For example, equation (\ref{bseq}) indicates that a particle under the influence of a background monopole is governed by the equation
\begin{equation*}-\nabla^2\psi + \frac{H^2+W^2-1}{r^2}\psi = \om^2\psi,\end{equation*}
which is influenced by attractive $1/r$ potential. We also see the coupling of two kinds of particles. From equation (\ref{syseq}), we expect that two particles can interact according to the equation
\begin{equation*}\Big(-\nabla^2\begin{pmatrix}1&0\\0&1\end{pmatrix} + \begin{pmatrix}\frac{H^2+W^2-1}{r^2} &0\\0&\frac{2W^2}{r^2}\end{pmatrix} + (\vec{A}(x)\cdot \vec{L})\begin{pmatrix}0&1\\1&0\end{pmatrix}\Big)\begin{pmatrix}\psi\\\eta\end{pmatrix} = \om^2\begin{pmatrix}\psi\\\eta\end{pmatrix},\end{equation*}
where $\vec{A}(x) = \frac{2W}{r^2}\hat{z}$. The operator $\hat{Q}=(\vec{A}(x)\cdot \vec{L})\begin{pmatrix}0&1\\1&0\end{pmatrix}$ should be interpreted as a generalized angular momentum in this system. When the state is a eigenfunction of $\hat{Q}$ of eigenvalue 0, the system decouples.

The Dirac-type operator $\slashed{D}$ and $\slashed{D}^\dagger$ anti-commutes with $(-1)^{|\vec{L}|}$. Therefore, both $\slashed{D}^\dagger\slashed{D}$ and $\slashed{D}\slashed{D}^\dagger$ commutes with $(-1)^{|\vec{L}|}$ and so the system given by equations (\ref{DDBPS1}) and (\ref{DDBPS2}) decomposes naturally into two subsystems, of even or odd parity corresponding to $\vec{L}$. Moreover, since $\slashed{D}\slashed{D}^\dagger$ commutes with $(-1)^{|\vec{s}|}$ as well, the system of equation (\ref{DDBPS2}) is divided into the real part $(s=0)$ and the imaginary part $(s=1)$ and we get 4 subsystems. Due to the background gauge condition, the real part is being ignored and we only consider the imaginary part in this paper. It is precisely the communtativy that allows one to ignore completely the effect of the real part.

In this setup, the operator $(\hat{x}\cdot\vec{t})^2$ is interpreted as the electric charge. Our calculation showed that the charge can only be $0$ or $\pm 1$. It is due to the fact that we used the adjoint representation of $SU(2)$ and therefore we can consider the isospin part as a spin-1 particle. In fact, there is irreducible representations of $SU(2)$ which correspond to a spin-1/2 particle. Applying it in our setup, we can get states with generalized momentum equals half integar. Furthermore, to allow the possibility of higher electric charge, one may try to extend this setup to $SU(N)$ for $N>2$. In principle, there will be no problem carrying this procedure to the $SU(N)$ case except the fact that we do not have analytic solution of BPS monopoles. The analytic solution is of no importance in
our analysis and the isovector in the $SU(N)$ does give higher electric charge. It will be interesting to see how it changed the properties of the particles.

%In \cite{main}, they investigated the scattering states of the equation (\ref{DDBPS2}). However, the scattering states are from the real part of the system and therefore are considered to be not physical. The physical states are either boson self-interaction (with potential non-vanishing at infinity) or a system consisting of a massive particle coupled to a massless particle. It is not the case when $j\geq 1$ and there are scattering states that are physical. For example, in equation (\ref{j12}), we can let $c_4=c_6=c_{10}$ and the equation of $c_5$ essentially corresponds to the scattering states when $l=0$. It is expected that scattering states of higher angular momentum can exist so as to compare with the result calculated by adiabatic quantization in \cite{others1,others2}.

As noted previously, investigation of the perturbations of Yang-Mills equation does give interesting results. From the calculation above, one obtains more information of the perturbations. We hope that these information are useful to future research in studying monopoles arising in supersymmetric Yang-Mills field theory and higher gauge field theories.
\vskip0.2in \ \\
\textbf{Acknowledgement} The author thanks Siye Wu for helpful dicussions and comments and Tang Sin Ting for the advice on the numerical analysis. This work is supported in part by the University Grants Council (UGC) of Hong Kong, Competitive Earmarked Research Grants (CERG), No.~HKU706010P.

\end{document}